\documentclass[3p,times,twocolumn]{elsarticle}
 \biboptions{comma,sort&compress}
 
\usepackage{graphicx}
\usepackage{here}
\usepackage{ecrc}

\volume{00}

\firstpage{1}

\journalname{Nuclear and Particle Physics Proceedings}

\runauth{}

\jid{nppp}

\jnltitlelogo{Nuclear and Particle Physics Proceedings}

\usepackage{amssymb}

\usepackage[figuresright]{rotating}

\begin{document}

\begin{frontmatter}

\title{The New Charm-Strange Resonances in the 
$D^- K^+$ Channel}
 \cortext[cor0]{Talk given at 23rd High-Energy Physics 
 International Conference in Quantum Chromodynamics (QCD 20, 
 35th anniversary), 27-30 October 2020, Montpellier-FR}
\author[label1]{R.M. Albuquerque\fnref{fn1}}
\fntext[fn1]{Speaker.}
\ead{raphael.albuquerque@uerj.br}
\address[label1]{Faculty of Technology, Rio de Janeiro 
State University (FAT,UERJ), Brazil.}
\author[label2,label3]{S. Narison}
\ead{snarison@yahoo.fr}
\address[label2]{Laboratoire Univers et Particules 
de Montpellier, CNRS-IN2P3, Case 070, Place Eug\`ene
Bataillon, 34095 - Montpellier, France.}
\address[label3]{Institute of High-Energy Physics of 
Madagascar (iHEPMAD), University of Ankatso, 
Antananarivo 101, Madagascar.}
\author[label3]{D. Rabetiarivony}
\ead{rd.bidds@gmail.com}
\author[label3]{G. Randriamanatrika}
\ead{artesgaetan@gmail.com}

\pagestyle{myheadings}
\markright{ }
\begin{abstract}
We evaluate the masses and decay constants of the $0^+$ and 
$1^-$ open-charm $(\bar{c}\bar{d})(us)$ tetraquarks and 
molecular states from QCD spectral sum rules (QSSR) by using 
QCD Laplace sum rule (LSR). This method takes into account the 
stability criteria where the factorised perturbative NLO 
corrections and the contributions of quark and gluon 
condensates up to dimension-6 in the OPE are included. 
We confront our results with the $D^- K^+$ invariant mass 
recently reported by LHCb from $B^+ \to D^+(D^- K^+)$ decays. 
We expect that the resonance near the $D^- K^+$ threshold can 
be originated from the $0^{+}(D^-K^+)$ molecule and/or 
$D^- K^+$ scattering. 
The $X_0(2900)$ scalar state and the resonance $X_J(3150)$ 
(if $J = 0$) can emerge from a minimal mixing model, with a tiny
mixing angle $\theta_0 \simeq (5.2 \pm 1.9)^0$, between a 
scalar Tetramole $({\cal T}_{\!\!{\cal M}0})$ (superposition 
of nearly degenerated hypothetical molecules and compact 
tetraquarks states with the same quantum numbers), having a 
mass $M_{{\cal T}_{\!\!{\cal M}0}} = 2743(18)$ MeV, and the 
first radial excitation of the $D^- K^+$ molecule with mass
$M_{(DK)_1} = 3678(310)$ MeV. In an analogous way, the 
$X_1(2900)$ and the $X_J(3350)$ (if $J = 1$) could be a 
mixture between the vector Tetramole 
$({\cal T}_{\!\!{\cal M}1})$, with a mass 
$M_{{\cal T}_{\!\!{\cal M}1}} = 2656(20)$ MeV, and its first
radial excitation having a mass 
$M_{{\cal T}_{\!\!{\cal M}1}} = 4592(141)$ MeV with an angle 
$\theta_0 \simeq (9.1 \pm 0.6)^0$. A (non)-confirmation of 
these statements requires experimental findings of the 
quantum numbers of the resonances at $3150$ and $3350$ MeV.
\end{abstract}

\end{frontmatter}

\section{Introduction}
In this work, based on the paper in Ref.\,\cite{DK}, we 
attempt to estimate, from LSR, the masses and couplings of 
the $0^{+}$ and $1^-$ molecules and compact tetraquarks 
states for interpreting the recent LHCb data from 
$B\to D^+(D^-K^+)$ decays \cite{LHCb1,LHCb2}, where one finds 
two prominent peaks (units of MeV):
\begin{eqnarray*}
    \hspace*{-0.6cm} M_{X_0}(0^+) =
    (2866.3\pm 6.5\pm 2.0), 
    && \hspace{-0.45cm} \Gamma_{X_0} = (57.2\pm 12.9), 
    \\
    \hspace*{-0.6cm} M_{X_1}(1^-) = (2904.1\pm4.8\pm 1.3),
    && \hspace{-0.45cm} \Gamma_{X_1} = (110.3\pm 11.5). 
\end{eqnarray*}

We have studied in Ref.\,\cite{X5568} the masses and couplings 
of the $D^0K^0 (0^{+})$ molecule and of the corresponding 
tetraquark states decaying into $D^0K^0$ but not into 
$D^-K^+$ and we found the lowest ground state masses:
\begin{eqnarray*}
  M_{DK} = 2402(42)~{\rm MeV}, && \hspace{-0.45cm}
  f_{DK}=254(48)~{\rm keV},\\
  M_{\bar c\bar d{us}} = 2395(68)~{\rm MeV},&& \hspace{-0.45cm}
  f_{\bar c\bar d{us}}=221(47)~{\rm keV}.
\label{eq:lowest-dk}
\end{eqnarray*}
We have used this result to interpret the nature of the 
$D^*_{s0}$(2317) compiled by PDG\,\cite{PDG} where the 
existence of a $DK$ pole at this energy has been recently 
confirmed from lattice calculations of scattering 
amplitudes\,\cite{LATT}.

\begin{table*}[hbt]
\setlength{\tabcolsep}{1.5pc}
\newlength{\digitwidth} \settowidth{\digitwidth}{\rm 0}
\catcode`?=\active \def?{\kern\digitwidth}
\caption{Interpolating operators describing the scalar 
$(0^+)$ and vector $(1^-)$ molecules and tetraquark states.}
\label{tab:current}
\begin{tabular*}{\textwidth}{@{}l@{\extracolsep{\fill}}l}
\hline
\hline
Scalar states ($0^+$) & Vector states ($1^-$) \\
\hline
  Tetraquarks \\
 $ {\cal O}^0_{SS} = \epsilon_{i j k} \:\epsilon_{m n k} \left(
  u_i^T\, C \gamma_5 \,d_j \right) \left( \bar{c}_m\, 
  \gamma_5 C \,\bar{s}_n^T\right) $ &  ${\cal O}^1_{AP} = \epsilon_{m n k}\: \epsilon_{i j k}\left( \bar{c}_m\, \gamma_\mu C \,
  \bar{s}_n^T\right) \left(
  u_i^T\, C \,d_j \right) $ \\
  
$   {\cal O}^0_{PP} = \epsilon_{i j k} \:\epsilon_{m n k} \left(
  u_i^T\, C \,d_j \right) \left( \bar{c}_m\, 
  C \,\bar{s}_n^T\right) $ & ${\cal O}^1_{PA} = \epsilon_{m n k}\: \epsilon_{i j k}\left( \bar{c}_m\,  C \,
  \bar{s}_n^T\right) \left(
  u_i^T\, C\gamma_\mu \,d_j \right) $  \\
  
 $  {\cal O}^0_{VV} = \epsilon_{i j k} \:\epsilon_{m n k} \left(
  u_i^T\, C \gamma_5 \gamma_\mu \,d_j \right) \left( \bar{c}_m\, 
  \gamma^\mu \gamma_5 C \,\bar{s}_n^T\right) $ & $ {\cal O}^1_{SV} = \epsilon_{i j k} \:\epsilon_{m n k} \left(
  u_i^T\, C \gamma_5 \,d_j \right) \left( \bar{c}_m\, 
  \gamma_\mu \gamma_5 C \,
  \bar{s}_n^T\right) $ \\
  
 $  {\cal O}^0_{AA} = \epsilon_{i j k} \:\epsilon_{m n k} \left(
  u_i^T\, C \gamma_\mu \,d_j \right) \left( \bar{c}_m\, 
  \gamma^\mu C \,\bar{s}_n^T\right) $ & $ {\cal O}^1_{VS} =\epsilon_{i j k} \:\epsilon_{m n k} \left(
  u_i^T\, C \gamma_5 \gamma_\mu \,d_j \right) \left( \bar{c}_m\, 
  \gamma_5 C \,
  \bar{s}_n^T\right) $\\ \\
  Molecules \\
$  {\cal O}^0_{DK}=(\bar c\gamma_5 d)(\bar s\gamma_5 u)$& ${\cal O}^1_{D_1 K} =\left(\bar{c} \gamma_\mu \gamma_5 d \right) 
  \left( \bar{s} \gamma_5\,u\right) $  \\
${\cal O}^0_{D^*K^*}=(\bar c\gamma^\mu d)(\bar s\gamma_\mu u)$&$ {\cal O}^1_{D K_1} = \left(\bar{c} \gamma_5 d \right) 
  \left( \bar{s}\, \gamma_\mu \gamma_5 \,u\right) $\\
$ {\cal O}^0_{D_1K_1}=(\bar c\gamma^\mu\gamma_5 d)(\bar s\gamma_
  \mu\gamma_5 u)$ & $ {\cal O}^1_{D^* K^*_0} = \left(\bar{c} \gamma_\mu  d \right) 
  \left( \bar{s}\,u\right) $ \\
 $  {\cal O}^0_{D^*_0K^*_0}=(\bar c d)(\bar s u)$ & 
$ {\cal O}^1_{D^*_0 K^*} =   \left(\bar{c} \,d \right) 
  \left( \bar{s} \,\gamma_\mu \,u\right)$ \\
  \hline\hline
\end{tabular*}
\end{table*}

For the molecular state, we can interchange the $u$ and $d$ 
quarks in the interpolating current and deduce from SU(2) 
symmetry that the $D^-K^+(0^{+})$ molecule mass is 
degenerated with the $D^0K^0$ one. Compared with the LHCb data, 
one may invoke that this charged molecule can be responsible 
of the bump near the DK threshold around 2.4 GeV but is too 
light to explain the $X_{0,1}$ peaks.

For the tetraquark state, one may not use a simple SU(2)
symmetry (rotation of $u$ and $d$ quarks) to deduce the ones
decaying into $D^-K^+$ due to our present ignorance of the 
diquark dynamics (for some attempts see\,\cite{NEU1,NEU2}). 

Therefore, recent analysis based on QSSR at lowest order (LO) 
of perturbation theory (PT) using some specific tetraquarks 
and / or molecules configurations appear in the 
literature\,\cite{ZHANG,CHEN,WANG,STEELE} (see 
also\,\cite{AGAEV,TURC}) to explain these new candidates for 
the exotic states.

\section{The Laplace sum rule (LSR)}
We shall work with the Finite Energy version of the 
QCD Inverse Laplace sum rules (LSR) and their ratios
\cite{SVZa,SVZb,SNB1,SNB2,SNB3,SNB4,IOFFEb,RRY,DERAF,
BERTa,YNDB,PASC,DOSCH}:
\begin{eqnarray}
    \hspace*{-0.2cm} {\cal L}^c_n(\tau,\mu) &=&
    \int_{(M_c+m_s)^2}^{t_c}\hspace*{-0.5cm}dt~t^n~
    e^{-t\tau}\frac{1}{\pi} ~\mbox{Im}~\Pi_{\cal M,T}(t,\mu)~,
    \\ \nonumber \\
    {\cal R}^c_n(\tau) &=& \frac{{\cal L}^c_{n+1}} 
    {{\cal L}^c_n},
\label{eq:lsr}
\end{eqnarray}
where $M_c$ and $m_s$ are the on-shell / pole charm and 
running strange quark masses, $\tau$ is the LSR variable, 
$n=0,1$ is the degree of moments, $t_c$ is  the threshold 
of the ``QCD continuum" which parametrizes, from the
discontinuity of the Feynman diagrams, the spectral function 
is evaluated by the calculation of the scalar correlator 
defined as:
\begin{eqnarray}
    \hspace*{-0.6cm} \Pi_{\cal M,T}(q^2)\hspace*{-0.2cm} &=& 
    \hspace*{-0.25cm}i \hspace*{-0.15cm}
    \int \hspace*{-0.15cm}d^4x ~e^{-iqx}\langle 0\vert T 
    {\cal O}^J_{\cal M,T}(x) ~
    {\cal O}^{J ~\dagger}_{\cal M,T}(0) \vert 0\rangle~,
\label{eq:2-pseudo}
\end{eqnarray}
where ${\cal O}^{J}_{\cal M,T}(x)$ are the interpolating 
currents for the tetraquarks ${\cal T}$ and molecules 
${\cal M}$ states. The  superscript $J$ refers to the spin 
of the particles.  

\subsection{The Interpolating Operators}
We shall be concerned with the interpolating given in 
Table\,\ref{tab:current}. The lowest order (LO) perturbative
(PT) QCD expressions - including the quark and gluon condensates 
contributions up to dimension-six condensates of the 
corresponding two-point spectral functions - the NLO PT 
corrections, the QCD input parameters and further details 
of the QSSR calculations for those interpolating operators 
are given in the Ref.\cite{DK}.

\section{Tetraquarks and Molecules}
The sum rule analysis for the $0^{+}$ and $1^-$ states 
present similar features for all currents in Table\,
\ref{tab:current}. Then, we show only explicitly the analysis 
of the $SS$ tetraquark channel for a better understanding 
on the extraction of our results.

\subsection{$\tau$- and $t_c$-stabilities}
\begin{figure}[t]
\begin{center}
\begin{flushleft} {\bf a)} \end{flushleft}
\vspace{0.15cm}
\includegraphics[width=8cm]{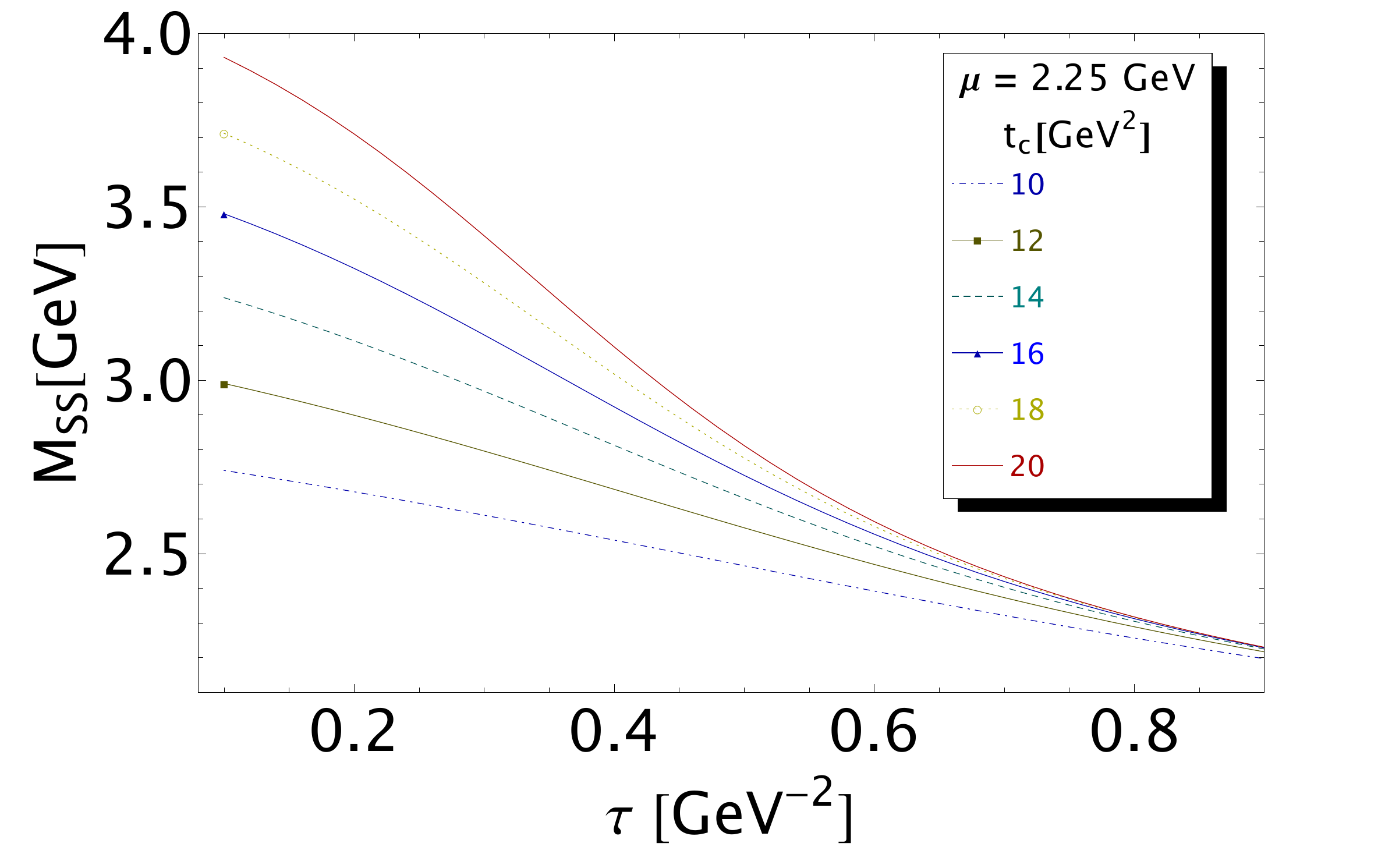}
\vspace*{-0.5cm}
\begin{flushleft} {\bf b)} \end{flushleft}
\includegraphics[width=8cm]{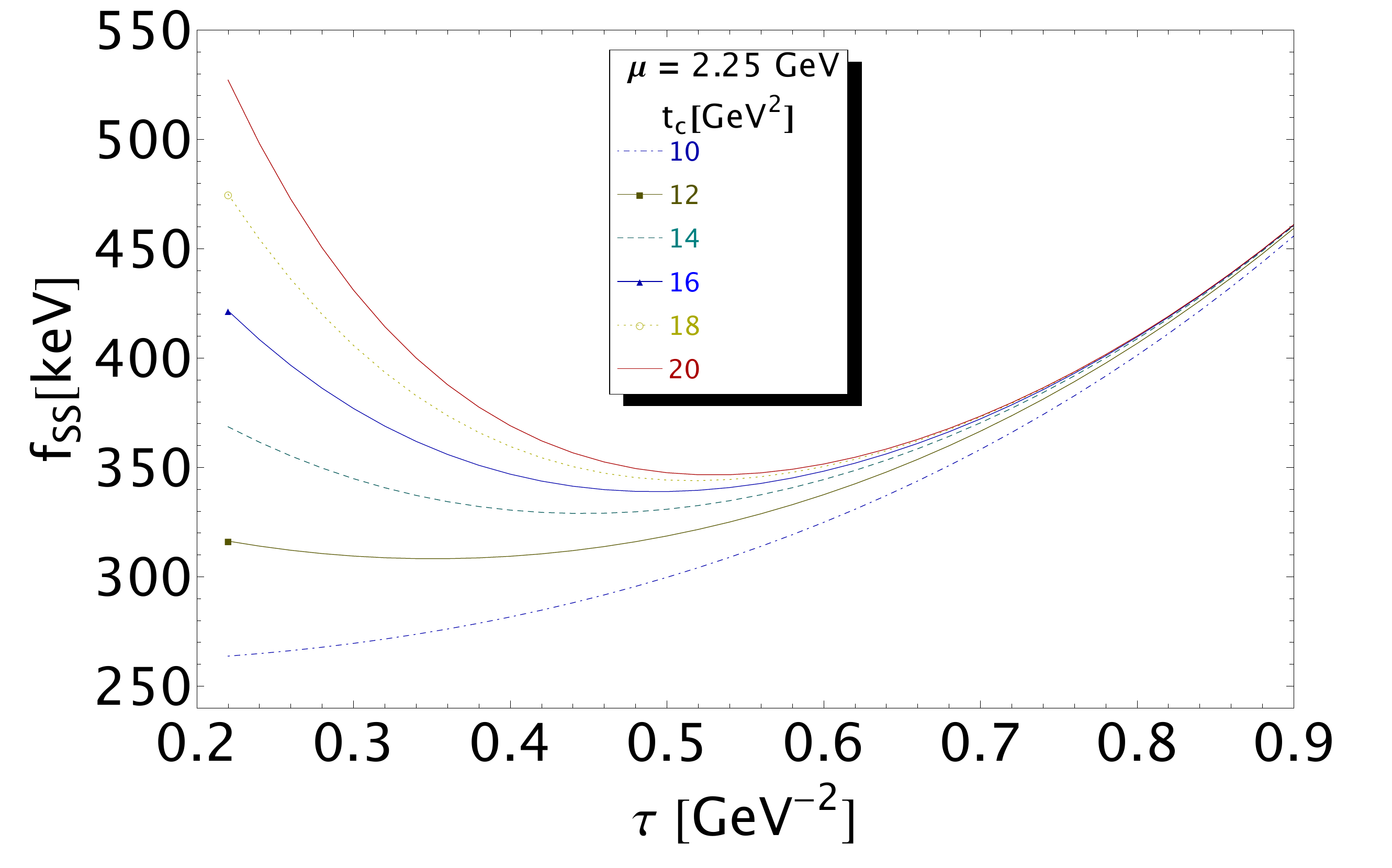}
\vspace*{-0.5cm}
\caption{\footnotesize  $f_{SS}$ and $M_{SS}$ as function of 
$\tau$ at NLO for different values of $t_c$, $\mu$=2.25 
GeV and the values of the QCD parameters given in 
Ref.\cite{DK}.} 
\label{fig:sstau}
\end{center}
\end{figure}

We show in Fig.\ref{fig:sstau}a) the $\tau$- and $t_c$- 
dependence of the mass obtained from ratio of moments 
${\cal R}_0$. The analysis of the coupling from the moment 
${\cal L}^c_0$ is shown in Fig.\,\ref{fig:sstau}b). The 
results stabilize at $\tau\simeq 0.5 $ GeV$^{-2}$
(inflexion point for the mass and minimum for the coupling). 

From Fig.\ref{fig:sstau}a), we extract the mass as the 
mean value of the one for $t_c\simeq$ 12 GeV$^2$ (beginning 
of the inflexion point) and of the one at beginning of 
$t_c$-stability of about 18 GeV$^2$. We use this (physical) 
mass value in ${\cal L}^c_0$ to draw Fig.\,\ref{fig:sstau}b). 
We check the range of $t_c$-values where the above-mentioned 
stabilities have been obtained by confronting 
Figs.\,\ref{fig:sstau}a) and b). Here, one can easily check 
that this range of $t_c$-values is the same for the mass and 
coupling. If the range does not coincide, we take the common
range of $t_c$ and redo the extraction of the mass. 

One can also see that the range of $\tau$-stabilities coincide 
in Fig.\,\ref{fig:sstau}a) (inflexion points) and in 
Fig.\,\ref{fig:sstau}b) (minimas). It is obvious that the 
value of $\tau$ from the minimum is more precise which we 
re-use to fix the final value of the mass. 

\subsection{$\mu$-stability}
\begin{figure}[hbt]
\begin{center}
\begin{flushleft} {\bf a)} \end{flushleft}
\vspace{0.15cm}
\includegraphics[width=8cm,height=5cm]{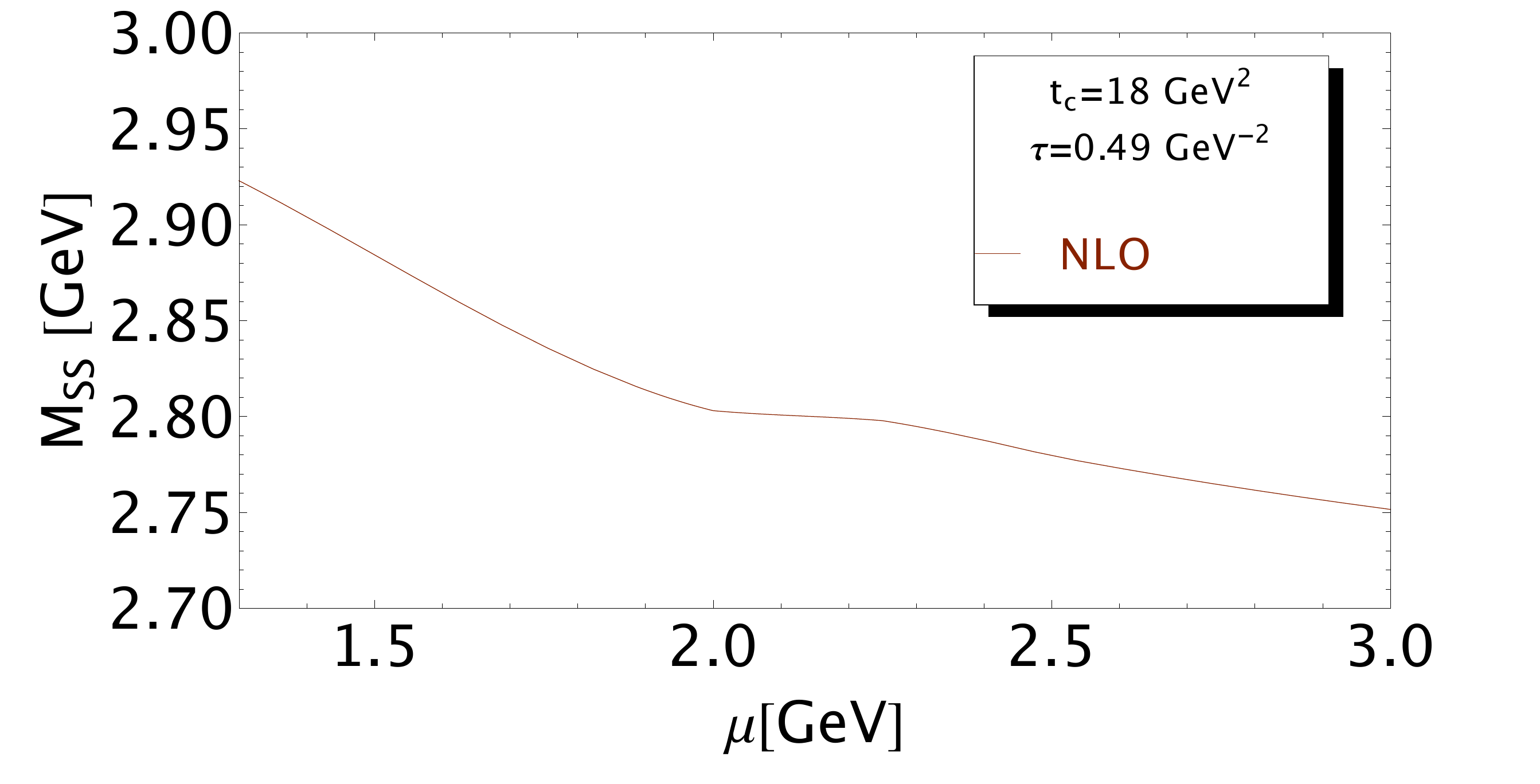}
\vspace*{-0.5cm}
\begin{flushleft} {\bf b)} \end{flushleft}
\includegraphics[width=8cm,height=5cm]{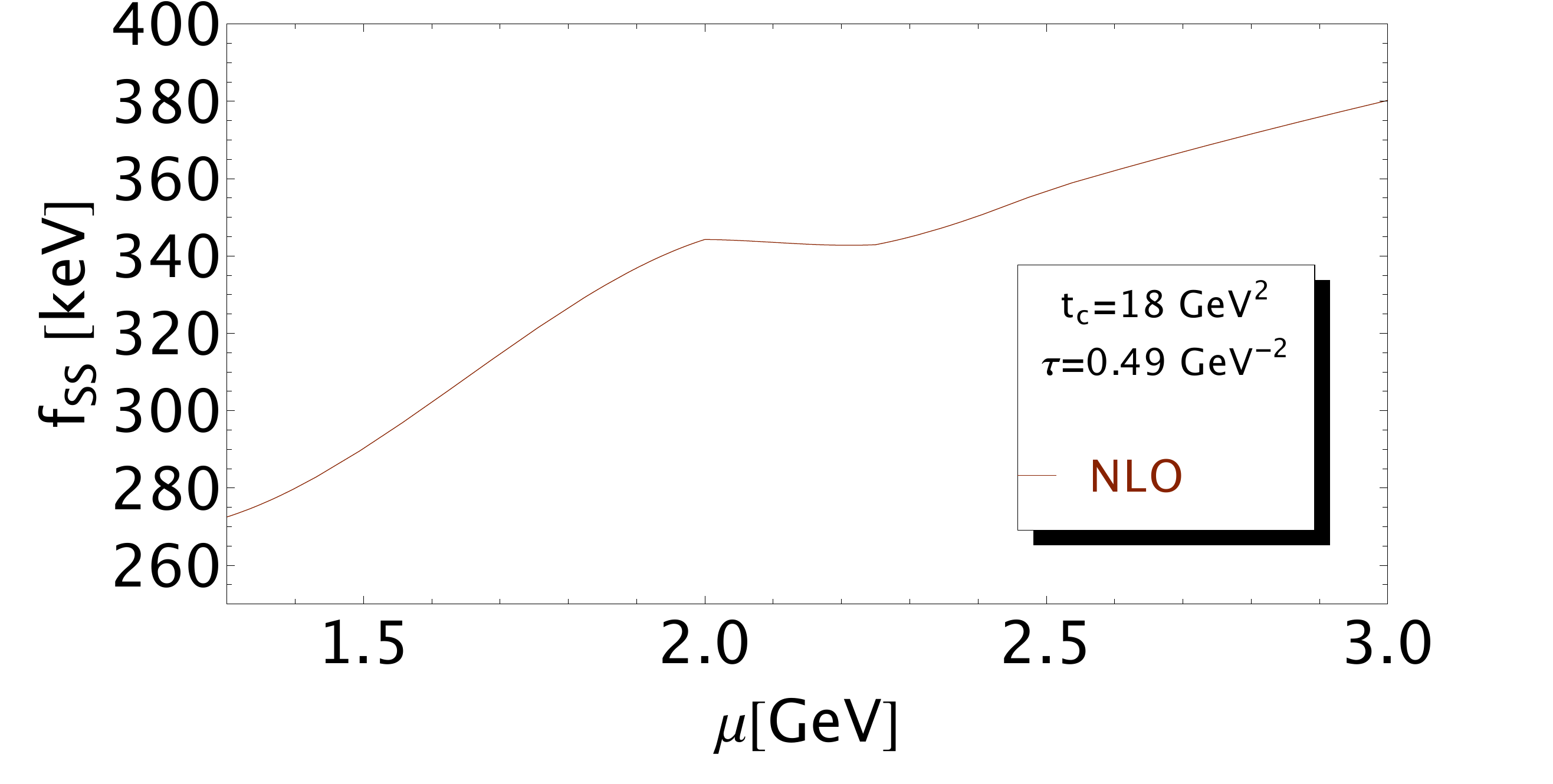}
\vspace*{-0.5cm}
\caption{\footnotesize $M_{SS}$ and $f_{SS}$ as function of 
$\mu$ at NLO for fixed values of $t_c$, $\tau$ and the
values of the QCD parameters given in Ref.\,\cite{DK}.} 
\label{fig:ssmu}
\end{center}
\end{figure}

In Fig.\,\ref{fig:ssmu}, we show the $\mu$-dependence of 
the results for given $t_c$=18 GeV$^2$ and $\tau$=0.49 
GeV$^{-2}$. One finds a common stability for 
$\mu = (2.25\pm 0.25)$ GeV, which confirms the result 
in Ref.\,\cite{X5568}.

\section{The First Radial Excitation}
We extend the analysis in Ref.\,\cite{X5568} by using a 
``Two resonances" + $\theta(t-t_c)$``QCD continuum" 
parametrization of the spectral function. To enhance the 
contribution of the 1st radial excitation, we shall also 
work with the ratio of moments ${\cal R}_1$ in addition 
to ${\cal R}_0$ for getting the masses. We use the same 
criteria involving the stability points in $(\tau, t_c,\mu)$ 
and the optimal results are given in Table\,\ref{tab:res}.
We observe that the mass-splittings between the first radial 
excitation and the lowest ground state are in order of 
$\sim 1500$ MeV, which is much bigger than 
$\sim 500 $ MeV typically used for ordinary mesons.

\section{Understanding LHCb Experimental Data}
Our results indicate that the molecules and tetraquark states
leading to the same final states are almost degenerated in 
masses. Therefore, we expect that the ``physical state" is a
combination of almost degenerated molecules and tetraquark 
states with the same quantum numbers $J^{PC}$ which we 
shall call:  {\it  Tetramole} $({\cal T_M}_J)$. 

\subsection{The $X_{0}(2866)$ and $X_{J}(3150)$ states}
Taking literally our results in Table\,\ref{tab:res}, one 
can see that we have three (almost) degenerate states:
\begin{eqnarray*}
    &&\!\!\!\!\!\!\!\!\!\!
    M_{SS}=2736(21)~{\rm MeV}, ~~~
    M_{AA}=2675(65)~{\rm MeV} ~~\\
    &&{\rm and}~~~~M_{D^*K^*}= 2808(41)~{\rm MeV}~,
\end{eqnarray*}
and their couplings to the corresponding currents are 
almost the same:
\begin{eqnarray*}
    &&\!\!\!\!\!\!\!
    f_{SS}=345(28)~{\rm keV},~~~~
    f_{AA}=498(43)~{\rm keV}, \\
    &&{\rm and}~~~~~f_{D^*K^*}=405(33)~{\rm keV}~,
\end{eqnarray*}
We assume that the physical state, hereafter called 
{\it Tetramole} $({\cal T_M}_J)$, is a superposition of 
these nearly degenerated hypothetical states having the same
quantum numbers. Taking its mass and coupling as (quadratic)
means of the previous numbers, we obtain:
\begin{eqnarray*}
    M_{ {\cal T_M}_0} \simeq  2743(18)~{\rm MeV}~,
    ~~~
    f_{ {\cal T_M}_0}\simeq  395(19)~{\rm keV} ~ 
\end{eqnarray*}
The ${\cal T_M}_0$ {\it tetramole} is a good candidate 
for explaining the $X_{0}(2866)$ though its mass is slightly
lighter. 

One can also see from Table\,\ref{tab:res} that the radial 
excitation $(DK)_1$ mass and coupling are :
\begin{eqnarray*}
    M_{(DK)_1} \simeq 3678(310)~{\rm MeV},
    ~~
    f_{(DK)_1} \simeq 199(62)~{\rm keV}
\end{eqnarray*}
which is the lightest $0^{+}$ first radial excitation. 
Assuming that the $X_{J}(3150)$ bump is a scalar state 
($J=0$), we attempt to use a {\it two-component minimal 
mixing model} between the {\it Tetramole} $({\cal T_M}_0)$ 
and the $(DK)_1$ radially excited molecule:
\begin{eqnarray*}
    \vert X_0(2866)\rangle &=& 
    \cos \theta_0 \vert {\cal T_M}_0 \rangle + 
    \sin\theta_0 \vert (DK)_1\rangle \\
    \vert X_{0}(3150)\rangle &=& 
    -\sin \theta_0 \vert  {\cal T_M}_0 \rangle + 
    \cos\theta_0 \vert (DK)_1\rangle~.
\end{eqnarray*}
We reproduce the data with a tiny mixing angle :
\begin{eqnarray*}
    \theta_0\simeq (5.2\pm 1.9)^0~.
\end{eqnarray*}

\subsection{The $X_{1}(2904)$ and $X_{J}(3350)$ states}
As one can see in Table\,\ref{tab:res}, there are 
four degenerate states with masses around $\sim 2650$ MeV,
and couplings around $\sim 200$ keV. We assume again that 
the (unmixed) physical state is a combination of these hypothetical states. We evaluate the mass and coupling of 
this {\it Tetramole} as the (geometric) means:
\begin{eqnarray*}
    M_{{\cal T_M}_1}=2656(20)~{\rm MeV},~~~~
    f_{{\cal T_M}_1}\simeq 229(12)~{\rm keV},
\end{eqnarray*}
where one may notice that it can contribute to the 
$X_{1}(2904)$ state but its mass is slightly lower. 
Looking at to the $1^-$ radial excitations in 
Table\,\ref{tab:res}, one can see that they are almost 
degenerated around 4.5 GeV from which one can extract
the masses and couplings (geometric mean) of the spin 1 
Tetramoles:
\begin{eqnarray*}
    \!\!\!\!\!\!\!
    M_{({\cal T_M}_1)_1} \simeq 4592(141)~{\rm MeV},~~~
    f_{({\cal T_M}_1)_1} \simeq 223(35)~{\rm keV}~.
\end{eqnarray*}
Then, we may consider a {\it minimal two-component mixing } 
of the spin 1 Tetramole ($ {\cal T_M}_1$) with its 1st radial
excitation $({\cal T_M}_1)_1$ to explain the $X_{1}(2904)$ 
state and the $X_{J}(3350)$ bump assuming that the latter is 
a spin 1 state. The data can be fitted with a tiny mixing 
angle :
\begin{eqnarray*}
    \theta_1\simeq (9.1\pm 0.6)^0~.
\end{eqnarray*}
A (non)-confirmation of these two {\it minimal mixing models}
requires an experimental identification of the quantum 
numbers of the bumps at 3150 and 3350 MeV.

\subsection{Final results}
Our final results are obtained at the stability points of the 
set of parameters $(\tau, t_c,\mu)$ and they are summarized 
in Table\,\ref{tab:res}. One can notice that, for some 
molecule and tetraquark states, the ground state mass values 
are above $5.5$ GeV which are too far to contribute to the 
LHCb observations in $DK$ invariant mass. In such cases, 
the sum rule results are discarded. One can find the full
analysis of different sources of errors, as well as an 
interesting discussion on the relevance of NLO calculations 
for sum rules in Ref.\cite{DK}.

\begin{table*}[hbt]
\setlength{\tabcolsep}{1.5pc}
\caption{LSR predictions, at NLO, for the decay constants and 
masses of the ground state ($f_0$, $M_0$), and their 
respective first radial excitation values ($f_1$, $M_1$), 
for the molecules and tetraquark states. 
The symbol ``$\ast$'' indicates that first radial excitation 
of high mass ground states were discarded in our sum 
rule analysis.}
\label{tab:res}
\begin{tabular*}{\textwidth}{@{}l@{\extracolsep{\fill}}
cccc}
\hline
\hline
Observables\, & $M_0$ (MeV) & $f_0$ (keV) 
& $M_1$ (MeV) & $f_1$ (keV)\\
\hline
{\bf \boldmath $0^+$ States}\\
\cline{0-0}
{\it Molecule}\\
${DK}$ & 2402(42) & 254(48) & 3678(310) & 211(51)\\
${D^*K^*}$ & 2808(41) & 405(33) & 4626(252) & 568(167)\\
${D_1K_1}$ & 5258(113) & 664(57) & $\ast$ & $\ast$\\
${D_0^*K_0^*}$ & 6270(160) & 249(18) & $\ast$ & $\ast$\\ \\
%
{\it Tetraquark}\\
${SS}$ & 2736(21) & 345(28) & 4586(268) & 359(81)\\
${AA}$ & 2675(65) & 498(43) & 4593(289) & 547(95)\\
${VV}$ & 5704(149) & 713(66) & $\ast$ & $\ast$\\
${PP}$ & 5917(98) & 538(41) & $\ast$ & $\ast$\\ \\
{\bf \boldmath $1^-$ States}\\
\cline{0-0}
{\it Molecule}\\
${D_1K}$ & 2676(47) & 191(21) & 4582(414) & 157(71)\\
${D_0^*K^*}$ & 2744(41) & 216(22) & 4662(269) & 237(63)\\
${DK_1}$ & 5377(166) & 351(31) & $\ast$ & $\ast$\\
${D^*K_0^*}$ & 5358(153) & 255(23) & $\ast$ & $\ast$\\ \\
{\it Tetraquark}\\
${PA}$ & 2666(32) & 285(29) & 4571(213) & 258(82)\\
${SV}$ & 2593(31) & 259(25) & 4541(345) & 243(68)\\
${AP}$ & 5542(139) & 416(38) & $\ast$ & $\ast$\\
${VS}$ & 5698(175) & 412(43) & $\ast$ & $\ast$\\ \\
\\
\hline\hline

\end{tabular*}

\end{table*}

\section{Summary and conclusions}
$\bullet~$ 
Motivated by the recent LHCb data on the $D^-K^+$ invariant 
mass from $B\to D^+D^-K^+$ decay, we have systematically
calculated the masses and couplings of some possible
configurations of the molecules and tetraquarks states using 
QCD Laplace sum rules (LSR) within stability criteria where 
we have added to the LO perturbative term, the NLO radiative
corrections, and the contributions from quark and 
gluon condensates up to dimension-6 in the OPE.

$\bullet~$ 
The peak around the $DK$ threshold can be due to $DK$ 
scattering amplitude $\oplus$ the $DK(2400)$ lowest mass 
molecule.

$\bullet~$ 
The ($0^{+}$) $X_0(2866)$ and $X_J(3150)$ (if it is a 
$0^{+}$ state) can e.g result from a mixing of the {\it 
Tetramole} ($ {\cal T_M}_0$) with the 1st radial excitation 
$(DK)_1$ of the molecule state $(DK)$ with a tiny mixing 
angle $\theta_0\simeq (5.2\pm 1.9)^0$.

$\bullet~$ 
The ($1^{-}$) $X_1(2904)$ and $X_J(3350)$ (if it is a 
$1^{-}$ state) can result from a mixing of the {\it Tetramole} 
($ {\cal T_M}_1$) with its 1st radial excitation 
$({\cal T_M})_1$  with a tiny mixing angle 
$\theta_1\simeq (9.1\pm 0.6)^0$.

$\bullet~$ 
More data on the precise quantum numbers of the 
$X_J(3150)$ and $X_J(3350)$ states are needed 
for testing the previous two {\it minimal mixing models} 
proposal.


\end{document}